\documentclass[superscriptaddress,prl,english,floatfix,twocolumn,10pt]{revtex4-2}
\usepackage{graphicx}
\graphicspath{figures/}
\usepackage{float}
\usepackage{amssymb}
\usepackage{amsmath}
\usepackage{mathtools}
\usepackage[utf8]{inputenc}
\usepackage{braket}
\usepackage{subfigure}
\usepackage{csquotes}
\usepackage{hhline}
\usepackage{xr}

\usepackage{dcolumn}
\usepackage{tabularx}
\usepackage[colorlinks=true,linkcolor=blue,citecolor=blue,urlcolor=blue]{hyperref}
\usepackage{longtable}
\usepackage{listings}
\usepackage{xcolor}
\usepackage[export]{adjustbox}
\newcolumntype{C}{>{\centering\arraybackslash}X}

\begin{document}

\title{Genus-protected higher-order topological phases}

\author{Shahroze Shahab}
\affiliation{Department of Physics and Astronomy, National Institute of Technology, Rourkela, Odisha-769008, India}

\author{Hui Liu}
\affiliation{Department of Physics, Stockholm University, AlbaNova University Center, 106 91 Stockholm, Sweden}

\author{Daniel Varjas}
\affiliation{IFW Dresden and W{\"u}rzburg-Dresden Cluster of Excellence ct.qmat, Helmholtzstrasse 20, 01069 Dresden, Germany}
\affiliation{Department of Theoretical Physics, Institute of Physics, Budapest University of Technology and Economics, Műegyetem rkp. 3., H-1111 Budapest, Hungary}

\author{Ion Cosma Fulga}
\affiliation{IFW Dresden and W{\"u}rzburg-Dresden Cluster of Excellence ct.qmat, Helmholtzstrasse 20, 01069 Dresden, Germany}

\begin{abstract}
Higher-order topological phases (HOTPs) feature protected gapless modes on boundaries of higher codimension, such as the corners or hinges of a crystal.
They are understood as being protected by lattice symmetries: 
If the latter are broken, it becomes possible to remove the boundary modes without closing the bulk gap.
In this work, we present construction schemes for HOTPs protected solely by the bulk gap, by fundamental symmetries, and by the global topology of the system shape (its genus, or number of holes), independent of any crystalline symmetries. 
As long as the fundamental local symmetries are preserved, the resulting boundary states cannot be removed by any purely-surface perturbation.
\end{abstract}
\maketitle

\section{Introduction}
\label{sec:intro}

Strong topological insulators and superconductors (TIs/TSCs) are characterized by a gapped bulk spectrum accompanied by robust gapless boundary modes \cite{RevModPhys.82.3045KaneHasan,RevModPhys.83.1057QiZhang}. 
These boundary states are protected by fundamental local symmetries: time-reversal, particle-hole, and chiral symmetry \cite{AltlandZirnbauer1997,Schnyder2008,Kitaev2009}. 
The possible gapped Hamiltonians can be systematically organized according to their symmetry content within the ten-fold way classification \cite{AltlandZirnbauer1997,Schnyder2008,Kitaev2009,Ryu_2010}.

A recent development is the discovery of higher-order topological phases (HOTPs) \cite{doi:10.1126/scienceBBHTaylor,PhysRevB.96.245115Taylor2017,doi:10.1126/sciadv.aat0346Schindler2018}, which extend this paradigm by hosting gapless excitations on boundaries of codimension $n>1$. 
In such systems, the full boundary need not be gapless.
For instance, a two-dimensional (2D) second-order topological insulator features a gapped bulk and gapped edges, but gapless, zero-dimensional topologically protected modes appear on the boundary, for instance at the corners \cite{PhysRevB.96.245115Taylor2017,doi:10.1126/scienceBBHTaylor}.
In three dimensions (3D), second-order phases host gapless 1D states forming closed loops on the 2D boundary \cite{doi:10.1126/sciadv.aat0346Schindler2018}, and third-order phases exhibit gapless 0D modes, e.g. at the corners \cite{doi:10.1126/scienceBBHTaylor,PhysRevB.96.245115Taylor2017}. 
More generally, the boundary of a $n^\text{th}$-order TI/TSC in $d$ dimensions is gapped except for a $(d-n)$-dimensional manifold that hosts protected gapless modes. 
A paradigmatic example is the Benalcazar-Bernevig-Hughes (BBH) model \cite{doi:10.1126/scienceBBHTaylor}, where sublattice symmetry together with rotation symmetry lead to robust, mid-gap corner states.

Lattice symmetries play a central role in the classification of HOTPs.
According to the classification of Refs.~\cite{BrouwerPhysRevB.97.205135,BrouwerPhysRevX.9.011012,BrouwerWiley2020}, the latter fall within two classes: intrinsic and extrinsic.
Intrinsic HOTPs are bulk topological phases which explicitly require lattice symmetries, e.g. mirror, rotation, or inversion for their protection.
Provided that the system globally respects these spatial symmetries, it is not possible to remove the gapless boundary modes without closing the bulk gap.
Extrinsic HOTPs, on the other hand, are neither bulk phases, nor do they require any lattice symmetry.
As explained in Ref.~\cite{BrouwerWiley2020}, they can be understood as being the result of strong topological phases appearing on the boundary, rather than in the bulk.
For example, if one of the facets of a 3D cube-shaped crystal enters a Chern insulating phase, chiral hinge modes will appear on the hinges bordering that facet \cite{BrouwerWiley2020}.

Expanding the family of possible HOTPs has been a subject of intense research in the last decade, resulting in generalizations to a wide range of systems.
These include aperiodic HOTPs protected by rotation symmetries that do not occur in crystals \cite{Fulga2019PhysRevLett.123.196401,Dong-Hui2020PhysRevLett.124.036803}, higher-order topological semimetals \cite{Taylor2018PhysRevB.98.241103,Ezawa2018PhysRevB.97.155305,Brouwer2022PhysRevB.106.035105,JianHua2020PhysRevLett.125.146401}, or systems with a so-called mixed-order topology \cite{Sur2022MixedOrder}.
Various systems have been put forward for their experimental realization, including elemental bismuth \cite{Schindler2018Bismuth}, SnTe \cite{doi:10.1126/sciadv.aat0346Schindler2018}, ${\mathrm{EuIn}}_{2}{\mathrm{As}}_{2}$ \cite{XiPhysRevLett.122.256402}, ${\mathrm{MnBi}}_{2n}{\mathrm{Te}}_{3n+1}$ \cite{PhysRevLett.124.136407}, as well as a wide range of meta-material platforms \cite{Xue2019MetaMaterial, Peterson2018MetaMaterial, SerraGarcia2018MetaMaterial, Wang2020JAP, PhysRevB.98.205147MetaMaterial, PhysRevB.100.201406MetaMaterial, PhysRevB.101.094107MetaMaterial, PhysRevB.109.134107MetaMaterial, PhysRevLett.122.204301MetaMaterial, PhysRevLett.122.233903MetaMaterial, PhysRevLett.126.146802MetaMaterial, PhysRevResearch.2.022028MetaMaterial}.
Despite this rapid progress and the large resulting variety of HOTPs, the intrinsic/extrinsic distinction has remained central to their classification.

Here, we explore a family of HOTP models that show a combination of intrinsic and extrinsic features.
On the one hand, their gapless boundary modes cannot be removed by any surface gap closing, provided that the fundamental local symmetries are preserved. 
On the other hand, these boundary modes do not require lattice symmetries for their protection. 
The combination of these features originates from the nontrivial global topology of the system and its boundary.
In this sense, our work parallels those which show how strong topological phases can show different properties when the system shape is nontrivial, e.g. in M\"{o}bius strip \cite{Hai2011MobiusPhysRevB.84.193106, Beugeling_2014} or Klein bottle geometries \cite{Yong2024KleinPhysRevB.109.134107,ravindran2025transition2dsymmetryprotected}.
Here, instead, we show that HOTPs can go beyond the intrinsic/extrinsic classification in systems with nonzero genus (or a nonzero number of holes).
Note that we use the word ``genus'' accurately in the case of 3D HOTPs, where it represents the number of handles of a closed 2D surface \cite{Hatcher2002}.
For the 2D systems we consider, however, the genus is strictly speaking zero, so we will use the letter $g$ and the term ``genus'' to refer to the number of holes present in the system (the number of boundary components minus one), as common in the physics literature.

\begin{figure}[tb]
\includegraphics[width=0.48\columnwidth, height=3.6cm]{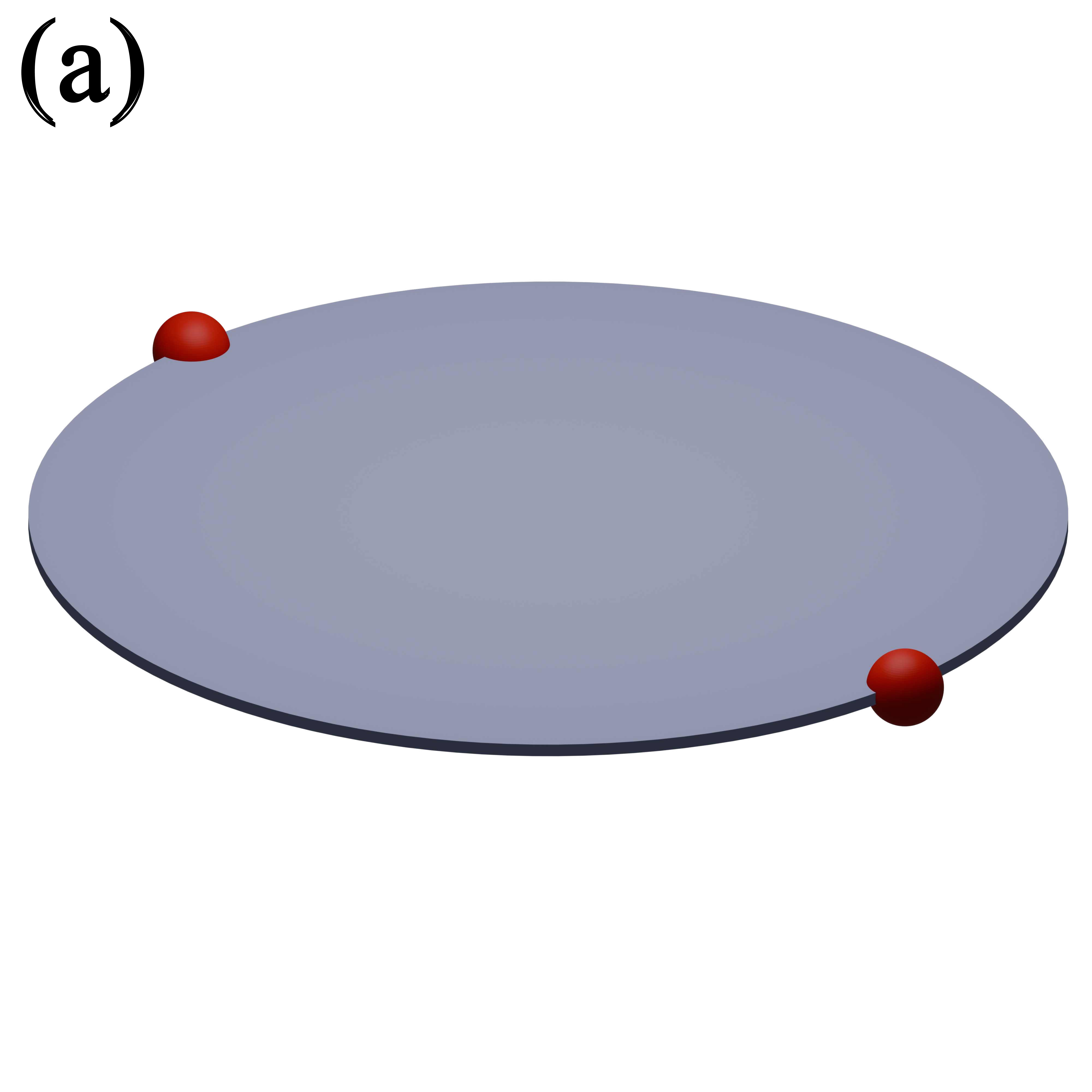}
\includegraphics[width=0.48\columnwidth, height=3.6cm]{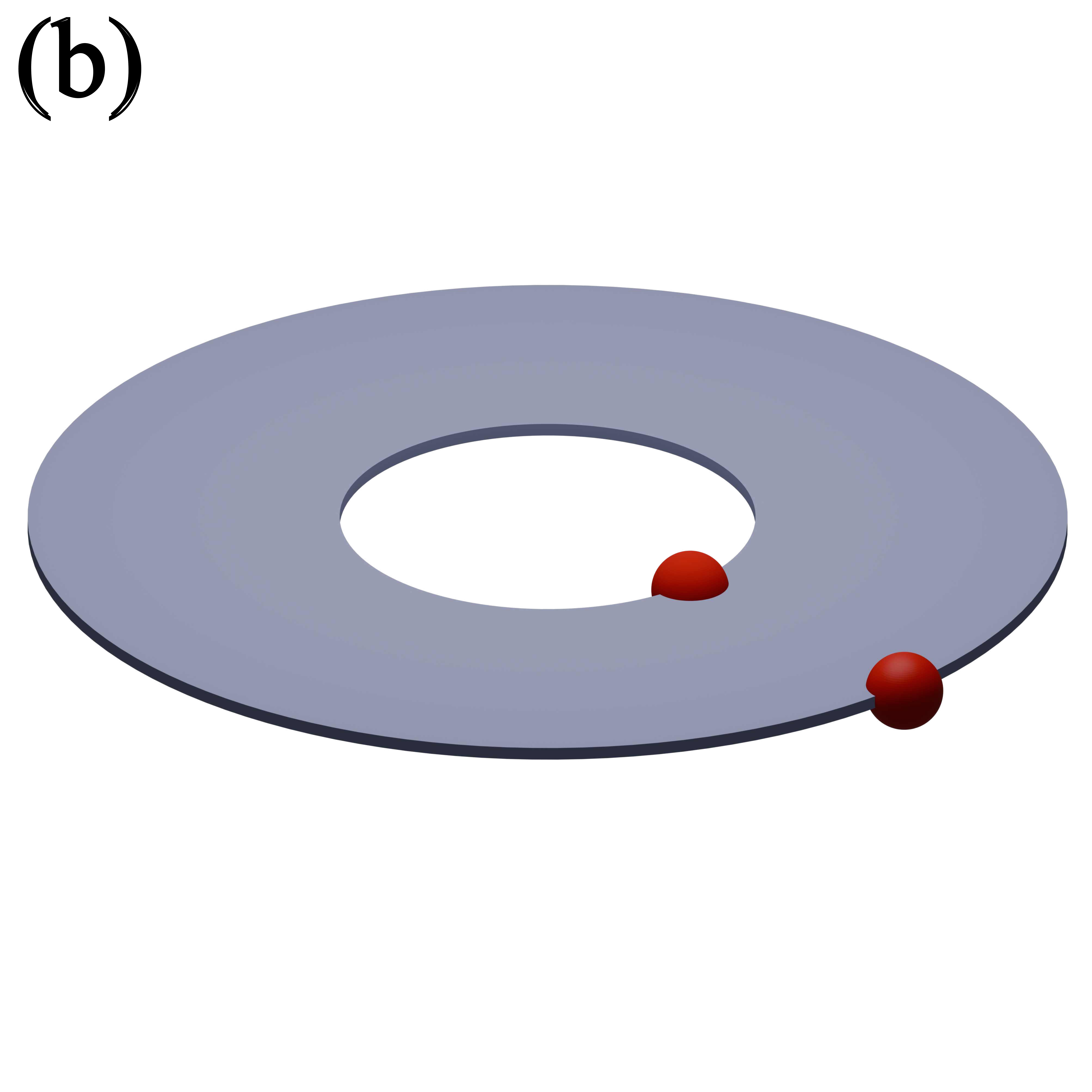}
\includegraphics[width=0.48\columnwidth, height=3.6cm]{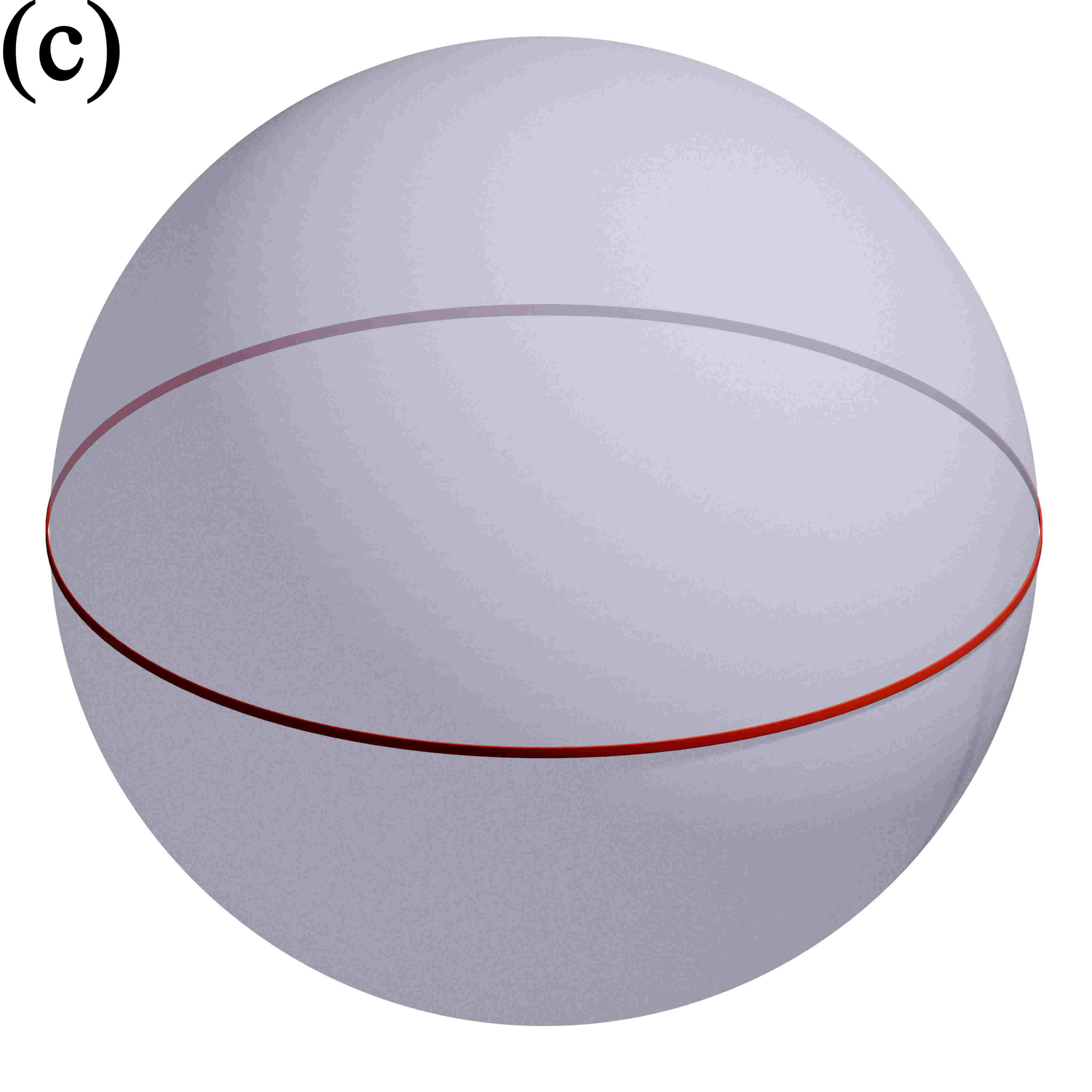}
\includegraphics[width=0.48\columnwidth, height=3.6cm]{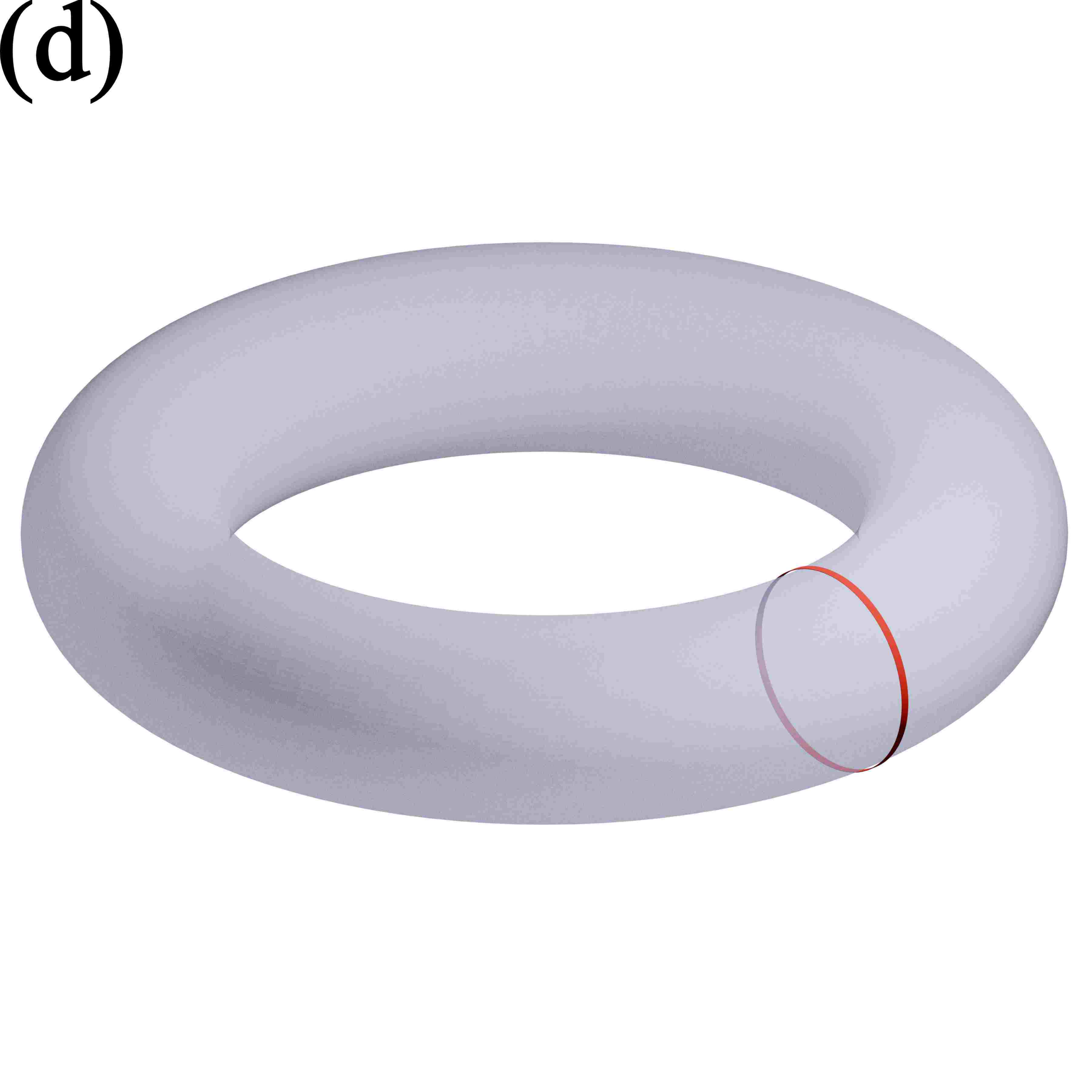}\\
\caption{
Second-order states on different geometries. 
(a) On the disk, zero modes can be gapped out by bringing them together along the edges. 
(b) In a Corbino-disk geometry, the zero modes cannot be removed without closing the bulk gap, simply because they occur on two disjoint edges. 
(c) On the sphere, the 1D edge mode can be moved along the surface until it shrinks to a point (e.g. at the north pole) and gaps out. 
(d) In a torus-shaped 3D system, propagating modes on non-contractible loops cannot be removed without traversing the bulk.
}
\label{fig:corbino_cartoon}
\end{figure}

A simple illustration of our main idea is shown in Fig.~\ref{fig:corbino_cartoon}.
Panel (a) shows a typical 2D second-order HOTP. 
Both the bulk and the edge are gapped, while two 0D gapless modes (red) appear at different points on the boundary.
For this topological phase to be protected by the bulk gap (as opposed to the surface gap), an additional lattice symmetry is required to prevent the modes from being moved along the boundary and pairwise annihilated.
Examples include inversion, a twofold rotation symmetry, or a mirror symmetry along the line connecting the two zero modes.
In contrast, no lattice symmetry is required in panel (b).
The Corbino disk has one hole ($g=1$), hosting two disjoint edges.
As such, if one zero mode forms on each edge, the two cannot be made to overlap by any perturbation that is restricted to the system boundary.

The same general principle applies also in higher-dimensional systems, as shown in panels (c, d) for the case of 3D HOTPs.
Genus $g=0$ systems (the sphere) require an additional lattice symmetry, lest the 1D topological mode be shifted along the surface until it contracts to a point and gaps out.
The 3D torus geometry (genus $g=1$), however, naturally features noncontractible loops.
A chiral edge mode formed along the red loop in Fig.~\ref{fig:corbino_cartoon}(d) cannot be gapped out without passing through the bulk, regardless of any symmetries.

In the following, we will show how genus-protected topological phases (GPT phases) can be constructed, focusing on concrete examples of HOTPs in 2D and 3D (Sec.~\ref{sec:method2d} and \ref{sec:method3d}), and confirming their lattice-symmetry-independent, higher-order topological nature by computing topological invariants based on the scattering matrix. All numerical calculations of spectra, real-space probability densities, and scattering-matrix invariants presented in this work were performed using the \texttt{Kwant} package~\cite{groth2014kwant}. In Sec.~\ref{sec:topological_classification} we will discuss their topological classification, in the sense of answering the following question:
For a 3D system with genus $g$ or for a 2D system with $g$ holes, and for a given set of fundamental symmetries, what is the total number of HOTPs that cannot be deformed into each other by purely boundary perturbations?
Finally, we conclude in Sec.~\ref{sec:conclusion}.

\section{Two-dimensional systems}
\label{sec:method2d}

Throughout this section, we will focus primarily on constructing HOTPs models based on two schemes: introducing local boundary defects and introducing global, topological defects.
Before we begin, however, we briefly point out the simplest example of a $g=1$ HOTP that we found, namely a cylinder composed of an odd number of Kitaev chains.

\begin{figure}[tb]
\centering
\includegraphics[width=0.4\columnwidth]{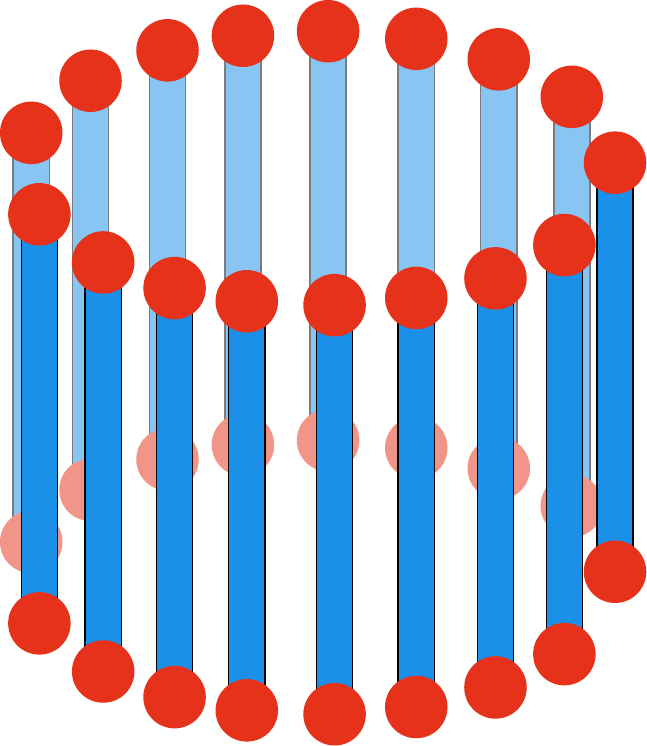}
\caption{
A cylinder consisting of an odd number of Kitaev chains (17 in this example). The 2D system is topologically equivalent to the Corbino disk of Fig.~\ref{fig:corbino_cartoon}(b), having one hole ($g=1$).
There are two disjoint edges, each of which hosts an odd number of Majorana modes.
}
\label{fig:kitaev_cylinder}
\end{figure}

The system is shown in Fig.~\ref{fig:kitaev_cylinder}.
It is two dimensional and its shape is topologically equivalent to that of the Corbino disk of Fig.~\ref{fig:corbino_cartoon}(b), having one hole ($g=1$).
Since the number of Majorana modes on each of its two disjoint edges is odd, this corresponds to a nontrivial $\mathbb{Z}_2$ topological invariant, as expected for 2D second-order HOTPs that belong to class D in the Altland-Zirnbauer classification \cite{AltlandZirnbauer1997}.
Provided that the total number of Kitaev chains forming the cylinder is odd, the coupling between neighboring chains can be arbitrary, provided it does not close the bulk gap or break particle-hole symmetry (PHS).
A PHS-preserving perturbation that is present only on the edge, e.g. adding dimerization on the edge, can be used to localize the edge modes at a particular position along the edge, or move them along the edge, similar to the scenario shown in Fig.~\ref{fig:corbino_cartoon}(b). 
However, the parity of the number of edge modes cannot be changed unless the bulk gap closes or PHS is broken.

\subsection{Local boundary defects}
\label{sec:bbh_model}

\begin{figure}[tb]
\centering
\includegraphics[width=\columnwidth]{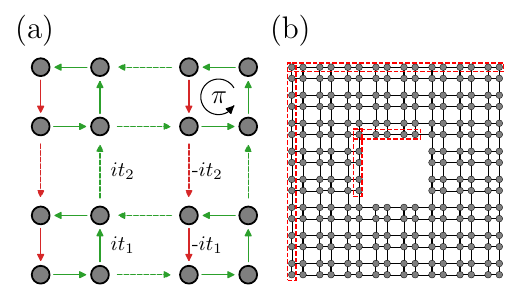}
\caption{Model of a 2D second-order topological superconductor Eq.~\eqref{eq:bbh_hamiltonian}. Gray circles denote Majorana modes. (a) Real-space representation of the model Hamiltonian with hopping amplitudes $t_{1,2}=t\pm\delta t$. Majoranas hopping along the red arrows acquire an additional $\pi$ phase, resulting in a magnetic flux threading each lattice plaquette. (b) Realization of a GPT phase by introducing local boundary defects on the square lattice with a square hole. Majorana modes located inside the red dashed rectangles are removed modifying the boundary geometry without affecting the bulk.}
\label{fig:bbh_lattice}
\end{figure}

\begin{figure}[tb]
\centering
\includegraphics[width=\columnwidth]{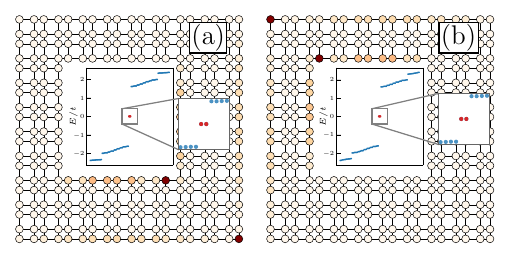}
\caption{
Probability density and low-energy spectrum near $E = 0$ of the Hamiltonian Eq.~\eqref{eq:bbh_hamiltonian} discretized on a Corbino-disk geometry with local edge defects, as illustrated in Fig.~\ref{fig:bbh_lattice}(b), for a system of size $(L,l) = (9.5,4.5)$. In the real-space plots, darker color indicates a higher probability density. The insets show the energy spectrum $E/t$ and the two red points correspond to the localized Majorana zero modes. 
(a) In the dimerized regime ($\delta t = -0.5t$), the system hosts two Majorana zero modes. 
(b) In the opposite regime ($\delta t = +0.5t$), two Majorana zero modes are again present, but they localize on the opposite pair of corners.
}
\label{fig:bbh_uniform}
\end{figure}

In this section, we construct a GPT phase by introducing local edge defects in a higher-order topological system. We consider the Benalcazar-Bernevig-Hughes (BBH) model~\cite{doi:10.1126/scienceBBHTaylor}, which hosts second-order topological modes protected by fourfold rotational symmetry. Majorana modes are placed on a square lattice and imaginary nearest-neighbor hopping is introduced, along with a magnetic flux of $\pi$ threading each plaquette. The model contains four Majorana modes in each unit cell, as illustrated in Fig.~\ref{fig:bbh_lattice}(a). The resulting momentum-space Hamiltonian is given by:
\begin{align}
H_{\text{D}}(\mathbf{k}) 
&= (t + \delta t)\,(\tau_3 \rho_2 + \tau_2 \rho_0) \nonumber \\
&\quad - (t - \delta t)\,\big[\cos(k_x)\,\tau_3 \rho_2 + \sin(k_x)\,\tau_3 \rho_1\big] \nonumber \\
&\quad - (t - \delta t)\,\big[\cos(k_y)\,\tau_2 \rho_0 + \sin(k_y)\,\tau_1 \rho_0\big],
\label{eq:bbh_hamiltonian}
\end{align}
where $t$ and $\delta t$ are real parameters and $\mathbf{k} = (k_x, k_y)$ denotes the crystal momentum. The parameter $\delta t$ represents a bond dimerization parameter that drives the system between trivial and higher-order topological phases. The matrices $\tau_i$ and $\rho_i$ ($i=1,2,3$) are independent Pauli matrices, with $\tau_0$ and $\rho_0$ denoting the identity matrices. 
For $\delta t < 0$, the model realizes a second-order topological superconductor on the square lattice, hosting Majorana corner states protected by PHS (realized by complex conjugation $\mathcal{K}$) and fourfold ($C_4$) rotation symmetry. However, the system resides in a trivial GPT phase as the corner Majorana modes can be brought together via purely edge perturbations and annihilated pairwise upon breaking the $C_4$ symmetry, without closing the bulk gap.

To construct a non-trivial GPT phase, we consider a finite sample of this lattice Hamiltonian on a square lattice with half a unit lattice constant, so the unit cell has unit length. The size of the full sample is $L$, and contains a square hole of side length $l$, this geometry is topologically equivalent to the Corbino disk shown in Fig.~\ref{fig:corbino_cartoon}(b). In particular, we choose the lattice termination such that the upper and left outer and inner edges terminate with half a unit cell, as illustrated in Fig.~\ref{fig:bbh_lattice}(b).
This is a purely boundary perturbation, leaving the bulk gap intact.
Note that even after removing the sites on the edges, the total number of Majorana modes in the lattice remains even, as expected for a tight-binding model of a topological superconductor.

\begin{figure}[tb]
\centering
\includegraphics[width=\columnwidth]{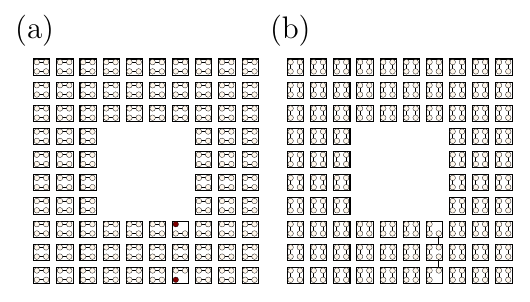}
\caption{BBH model with a non-uniform dimerization pattern. (a) The system hosts two MZMs in the topological regime, due to the presence of two local boundary defects (i.e. removing two sites). (b) Introducing a non-uniform dimerization pattern allows the realization of a trivial phase.}
\label{fig:bbh_nonuniform}
\end{figure}

We numerically simulate the model and plot the probability density and the spectrum of low-energy modes near $E=0$ in Fig.~\ref{fig:bbh_uniform}. In the dimerized limit ($\delta t < 0$), the system exhibits two localized corner modes, one on the inner edge and the other one on the outer edge [Fig.~\ref{fig:bbh_uniform}(a)]. Interestingly, we find that the system hosts localized corner modes even for $\delta t>0$ [Fig.~\ref{fig:bbh_uniform}(b)]. The system is thus in a GPT phase, in which the localized corner modes remain protected and cannot be brought together and annihilated, even upon breaking the $C_4$ symmetry, unless the bulk gap is closed or the underlying PHS is broken.
We show an example of how a trivial phase may be reached upon introducing a non-uniform dimerization pattern in Fig.~\ref{fig:bbh_nonuniform}.

\subsection{Topological defects}
\label{sec:global_defects}

Topological defects locally violate crystalline symmetries, while the rest of the lattice remains indistinguishable from a defect-free lattice. Such defects can be generated via a Volterra construction \cite{Volterra1907Elastic, Fulga2019Crystalline, Dominic2018Crystalline, Friedel2008Disclinations}, in which a section of the lattice is cut and re-glued. 
Dislocations and disclinations can host topologically-protected modes, as discussed in Refs.~\cite{TeoKane2010, Teo_2013, Geier_2021, Liu2021Photonic}.
These modes are localized either at the core of the defect, or, when the latter is not part of the lattice, on the boundaries of the system.
Here our approach will be to consider systems with a nonzero number of holes/genus, and place the topological defects in the holes and outside the lattice.
Thus, while the bulk remains locally uniform and indistinguishable from the defect-free lattice everywhere, zero modes are generated on the inside and outside edges, enabling the construction of a GPT phase.

\subsubsection{Dislocations}
\label{sec:dislocations}
A lattice dislocation is constructed by removing a line of sites and reconnecting the lattice across the cut such that the bulk remains uniform. We begin with the class-D Hamiltonian \cite{diez2014bimodal}:
\begin{equation}
\label{eq:dislocation_hamiltonian}
\begin{aligned}
H(\mathbf{k}) &= \epsilon(\mathbf{k}) \tau_{3} 
+ \Delta_{x} \tau_{1} \sin k_{x} 
+ \Delta_{y} \tau_{2} \sin k_{y}, \\
\epsilon(\mathbf{k}) &= -2t_{x} \cos k_{x} 
- 2t_{y} \cos k_{y} 
- \mu.
\end{aligned}
\end{equation}

\begin{figure}[tb]
\centering
\includegraphics{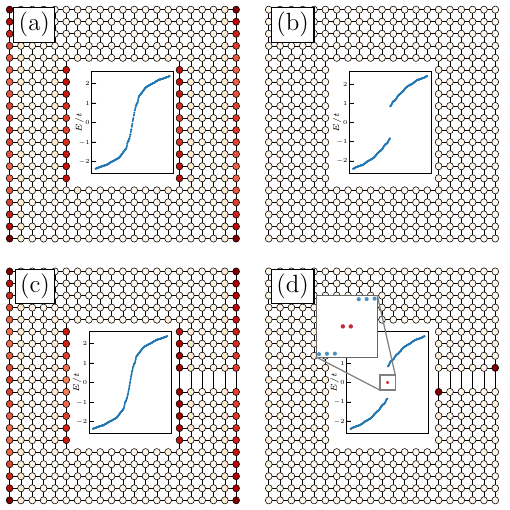}
\caption{Probability density and low-energy spectrum near $E=0$ of the class-D Hamiltonian Eq.~\eqref{eq:dislocation_hamiltonian} discretized on a Corbino-disk geometry, evaluated at $\alpha=0.5$, $\Delta=t$, $\mu=0.2t$ and system size $(L,l)=(21, 9)$, corresponding to the weak topological insulator (WTI) phase.
(a) Corbino geometry without a dislocation, hosting Majorana edge modes on both the inner and outer boundaries.
(b) Introducing an alternating weak-strong hopping pattern along the inner and outer edges allows the edge modes to be pairwise annihilated rendering the system topologically trivial.
(c) Introducing a dislocation by removing a row of sites renders the number of sites along both the inner and outer perimeters odd, placing the system in a GPT phase.
(d) Two localized Majorana modes remain along the inner and outer edges after applying the surface perturbation.
In the real-space plots, darker color indicates a higher probability density. The two red points correspond to the localized Majorana zero modes in panel (d).
}
\label{fig:kitaev_corbino}
\end{figure}

Here $(\Delta_x,\Delta_y) = (\Delta,\alpha \Delta)$ denotes the anisotropic amplitude of the chiral $p$-wave pair potential, $(t_x,t_y) = (t,\alpha t)$ is the anisotropic hopping amplitude, and $\mu$ is the chemical potential. 
We set $t$ as the energy scale of the model and express all other energies relative to it.
The parameter $\alpha \in [0,1]$ characterizes the degree of anisotropy. The Hamiltonian $H(\mathbf{k})$ possesses particle-hole symmetry implemented by $\tau_1 {\cal K}$
and realizes a weak topological insulator (WTI), whose $\mu$-$\alpha$ phase diagram is shown in Fig.~3 of Ref.~\cite{diez2014bimodal}. We discretize the Hamiltonian on a square lattice with unit lattice constant and size of $L$ unit cells, containing a square hole having a side length of $l$ unit cells as done previously in Sec.~\ref{sec:bbh_model}.  The energy spectrum and the probability density of modes near $E=0$ are computed for parameters corresponding to the WTI phase~\cite{diez2014bimodal} and shown in Fig.~\ref{fig:kitaev_corbino}. However, these modes can be pairwise annihilated by purely surface perturbations, rendering the system topologically trivial. This is explicitly demonstrated by introducing alternating weak and strong hoppings along the inner and outer edges [Fig.~\ref{fig:kitaev_corbino}(b)].

To introduce a dislocation, we remove a row of sites, rendering the number of sites along both the inner and outer perimeters odd [Fig.~\ref{fig:kitaev_corbino}(c,d)]. Two localized Majorana modes remain along the inner and outer edges after applying the edge perturbation as before. These modes cannot be annihilated without closing the bulk gap. The system therefore realizes a GPT phase [Fig.~\ref{fig:kitaev_corbino}(d)].
To demonstrate the topological nature of the system, we evaluate the reflection-matrix topological invariant \cite{Akhmerov_2011, Fulga2011Scattering, Fulga2012Scattering}. The reflection matrix, a subblock of the full scattering matrix $S$, provides direct access to the topological index. Since the Hamiltonian Eq.~\eqref{eq:dislocation_hamiltonian} belongs to class D, there exists a basis in which the reflection matrix $r$ is real.

\begin{figure}[tb]
\centering
\includegraphics[width=0.48\columnwidth, height=3.6cm]{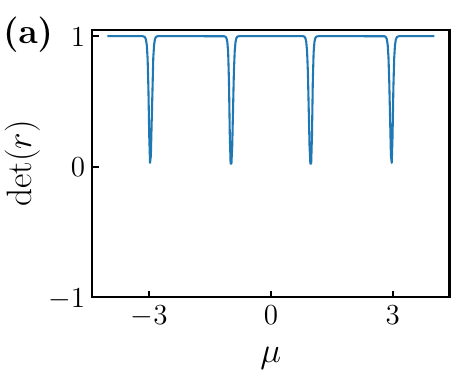}
\hfill
\includegraphics[width=0.48\columnwidth, height=3.6cm]{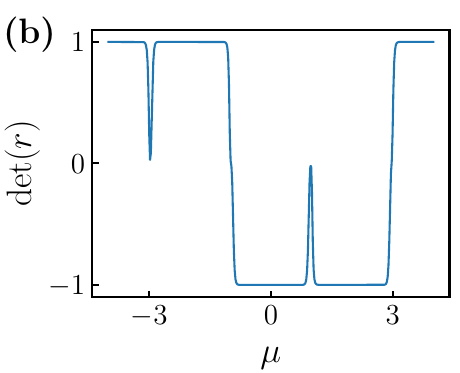}
\caption{
Reflection-matrix determinant $\det(r)$ for the class-D Hamiltonian Eq.~\eqref{eq:dislocation_hamiltonian} on a Corbino geometry with leads attached to the inner and outer perimeters, evaluated at $\alpha=0.5$, $\Delta=t$, $(L,l) = (40,10)$ and varying chemical potential $\mu$.
(a) Without a dislocation, where $\det(r)=+1$.
(b) With a dislocation, realizing the GPT phase, where $\det(r)$ changes sign.
The spikes in $\det(r)$ indicate bulk gap-closing points where the invariant becomes ill-defined.
}
\label{fig:dislocation_invariant}
\end{figure}

We consider a two-terminal geometry in which leads are attached to the inner and outer perimeters of the Corbino geometry. The real reflection matrix $r$, a diagonal sub-block of the full scattering matrix $S$, defines the topological invariant $\nu = \mathrm{sign}\det(r)$. When the bulk gap closes, $\det(r)=0$, rendering the invariant ill-defined. To illustrate this behavior, we compare two representative cases: a trivial phase without a dislocation and a GPT phase with a dislocation. The corresponding dependence of $\det(r)$ on the chemical potential $\mu$ is shown in Fig.~\ref{fig:dislocation_invariant}. In the trivial phase, $\det(r)$ remains fixed at $+1$, whereas in the GPT phase it changes sign and takes the value $-1$ over a finite parameter range, signaling a topological phase.

\subsubsection{Disclinations}
\label{sec:disclinations}

\begin{figure}[tb]
\centering
\includegraphics[width=\columnwidth, height=3.6cm]{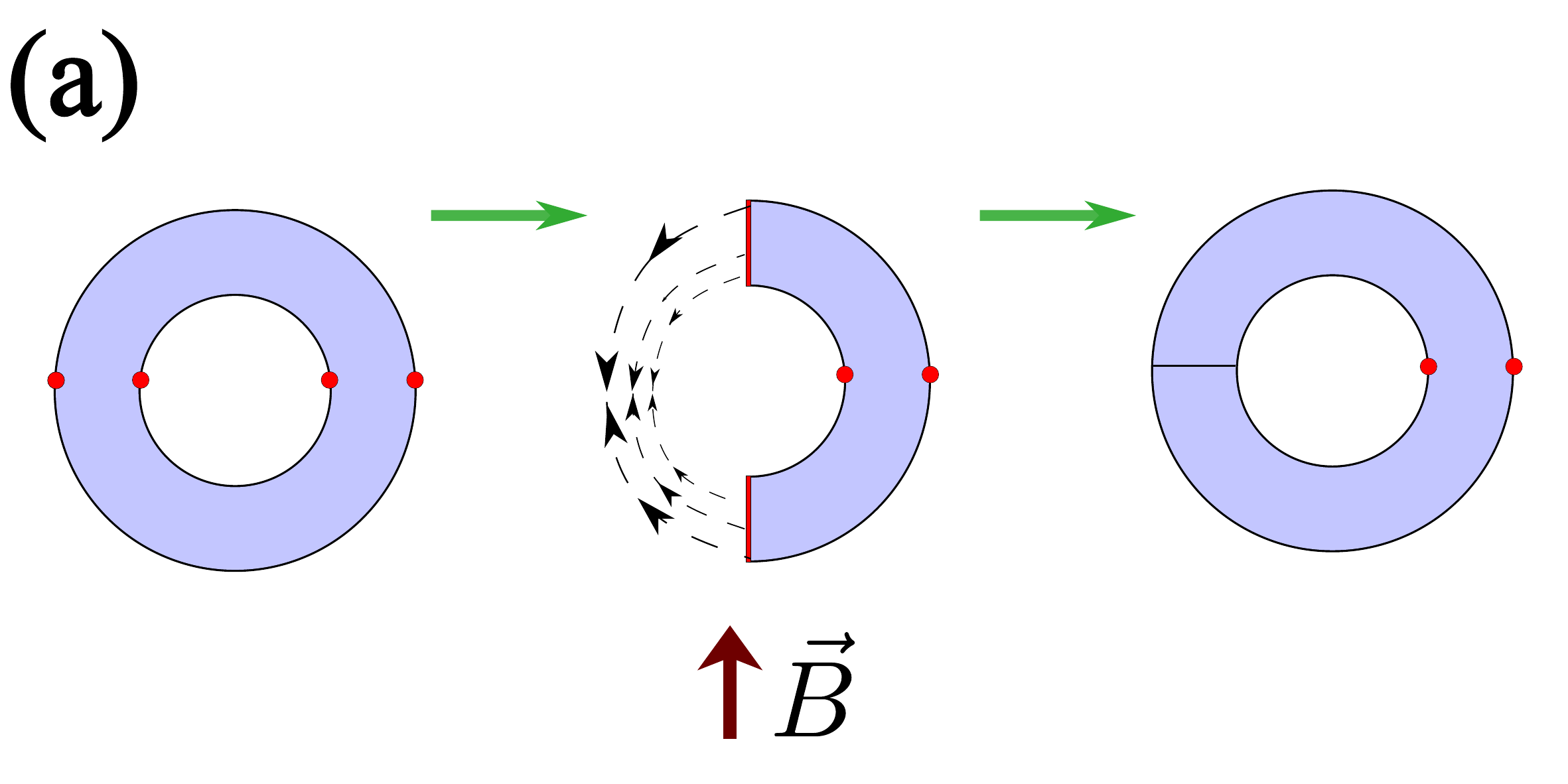}
\includegraphics[width=\columnwidth, height=3.6cm]{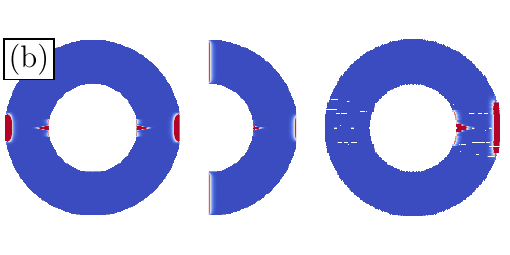}
\caption{(a) Volterra construction of a $180^\circ$ disclination on a Corbino disk. 
In the presence of an in-plane magnetic field $\vec{B}$, the defect-free geometry (top left) hosts four Majorana zero modes (red dots) along the edges. 
Cutting the disk and re-gluing after a $\pi$ rotation (right) modifies the hoppings across the branch cut, leaving only two unpaired MZMs. 
These zero modes are protected by the global topology of the lattice (i.e. its genus), unlike in the defect-free case where purely surface perturbations can cause them to annihilate pairwise.
(b) Real-space probability density of the low-energy modes of the Hamiltonian Eq.~\eqref{eq:qsh2d} on a Corbino geometry of size $(R,r) = (100,50)$ with a $180^\circ$ disclination, computed for an in-plane Zeeman field $B_0=1.5$ and parameters $A=1.0$, $B=1.0$, and $M=2$. The color scale represents the probability density, with red indicating regions of highest weight and blue corresponding to negligible weight. Two localized Majorana zero modes remain bound to the inner and outer edges after introducing the disclination, demonstrating the genus-protected topological phase.
}
\label{fig:disclination}
\end{figure}

Disclinations are topological lattice defects that can be generated through a Volterra construction: the lattice is cut along radial lines and the exposed edges are re-glued after a rotation. 
In our case, we focus on a $180^\circ$ disclination, where the cut edges are reattached after a $\pi$ rotation [Fig.~\ref{fig:disclination}]. 
This procedure leaves the bulk Hamiltonian unchanged everywhere except along the cut, where the hopping terms are modified. 
The disclination can thus be regarded as a domain wall separating regions that differ by a local order parameter \cite{Friedel2008Disclinations}.

To illustrate this, we begin with a minimal model of a quantum spin-Hall (QSH) \cite{Kane_2005, Bernevig_2006} insulator on a Corbino disk without any disclination. 
In momentum space, the Hamiltonian is
\begin{equation}
\begin{aligned}
H(\mathbf{k}) &= M(\mathbf{k}) \tau_z
 - A\bigl(\sin k_x \,\sigma_x - \sin k_y \,\sigma_y\bigr)\tau_x,\\
M(\mathbf{k}) &= M - B\bigl(4 - 2\cos k_x - 2\cos k_y\bigr),
\end{aligned}
\label{eq:qsh2d}
\end{equation}
where $\mathbf{k}$ is the crystal momentum and $\sigma$, $\tau$ act on spin and orbital degrees of freedom, respectively. 
In addition to time-reversal symmetry (TRS), implemented by $i\sigma_y \mathcal{K}$, the Hamiltonian $H(\mathbf{k})$ also possesses a PHS given by $\sigma_y \tau_y \mathcal{K}$.
For $0<M<4B$, the bulk realizes a QSH phase. 
On a Corbino geometry, this produces helical edge states localized at the inner and outer boundaries. 
Adding an in-plane Zeeman field, $\vec{B} = B_0 \,\hat{n}\cdot\sigma$, breaks the TRS and gaps these boundaries except at four points where $\hat{n}$ is perpendicular to the surface normal, generating four Majorana zero modes (MZMs), as shown in Fig.~\ref{fig:disclination}. 
Without further protection, these MZMs can be shifted along the edges and annihilated in pairs without closing the bulk gap implying that the system is in a trivial GPT phase.

\begin{figure}[tb]
\centering
\includegraphics[width=0.48\columnwidth, height=3.6cm]{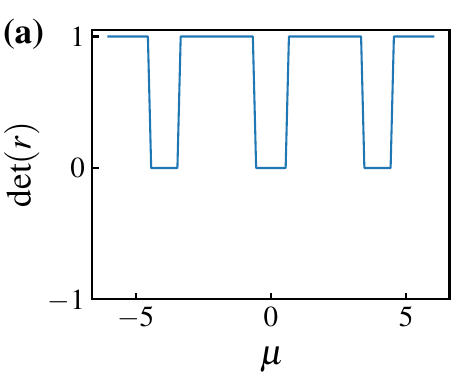}
\includegraphics[width=0.48\columnwidth, height=3.6cm]{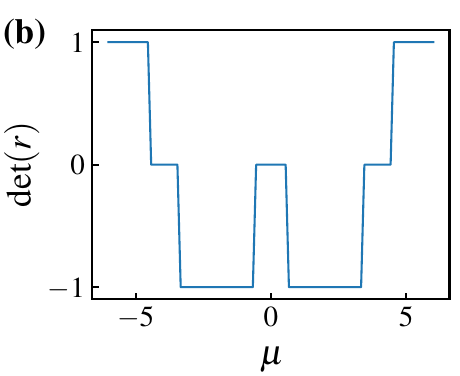}
\caption{
Reflection-matrix invariant $\det(r)$ for the QSH Hamiltonian Eq.~\eqref{eq:qsh2d} with an in-plane magnetic, $B_0 = 1.5$, $B=A=1.0$ on a Corbino geometry of size $(R,r) = (100,50)$ with leads attached to the inner and outer perimeters, and varying chemical potential $\mu = M-4B$.
(a) Without a dislocation, where $\det(r)=+1$.
(b) With a dislocation, realizing the genus-protected topological (GPT) phase, where $\det(r)$ changes sign.
The sign of $\det(r)$ distinguishes the two phases, while $\det(r) = 0$ indicates gapless regions where the invariant is ill-defined.
}
\label{fig:disclination_invariant}
\end{figure}

Introducing a $180^\circ$ disclination through a Volterra process alters this picture.
Following the construction detailed in Ref.~\cite{Geier_2021} we denote a $\pi$ rotation by $\mathcal{R}_{\pi}$, and identify the real-space positions $\mathbf{r}$ and $\mathcal{R}_{\pi}\mathbf{r}$ along the cut. The coordinate system and the local internal degrees of freedom of two adjacent unit cells across the cut are related by a relative rotation of $\pi$. A particle hopping across the cut must therefore respect this local change of basis and its wavefunction $\ket{\psi_{\mathbf{r}}}$ transforms as $U(\mathcal{R}_\pi)\ket{\psi_{\mathbf{r}+a_n}}$, where $U(\mathcal{R}_\pi)$ denotes the representation of a $\pi$ rotation acting on the internal degrees of freedom and $a_n$ is a lattice vector connecting the two adjacent cells. Since the points $\mathbf{r}$ and $\mathcal{R}_{\pi}\mathbf{r}$ are identified, the hopping across the cut must include this basis transformation and the hopping terms across the branch cut transforms as
\begin{equation}
H^{\text{cut}}_{r,\, r + a_n} = U(\mathcal{R}_\pi)\, H_{r,\, r + a_n},
\end{equation}
where $H_{r,\, r + a_n}$ is the bulk hopping matrix element between sites $r$ and $r+a_n$. This identification makes the hopping across the branch cut indistinguishable from the corresponding hopping in the bulk.
In our QSH model, the $\pi$ rotation acts on the internal degrees of freedom as
\begin{equation}\label{eq:URpi}
U(\mathcal{R}_\pi) = \sigma_0 \tau_z .
\end{equation} 
After adding this rotation to the hopping matrix element, two unpaired MZMs remain at the edges. The system is now in a GPT phase as the two MZMs cannot be brought together and annihilated without closing the bulk gap or breaking the underlying PHS of the Hamiltonian. In this way, the global topology of the lattice protects the zero modes, in contrast to the trivial GPT phase without disclination. To further substantiate our claim, we numerically simulate the Hamiltonian of Eq.~\eqref{eq:qsh2d} on a Corbino geometry with unit lattice constant, outer radius $R=100$, and inner radius $r=50$. We compute the probability density of the low-energy modes for parameters corresponding to the QSH phase of Hamiltonian Eq.~\eqref{eq:qsh2d} and display the results in Fig.~\ref{fig:disclination}(b).
We again employ the scattering-matrix invariant in a two-terminal geometry, introduced in Sec.~\ref{sec:dislocations}, to verify that the system realizes a GPT phase. Since the QSH model considered here possesses an additional PHS and TRS is explicitly broken by the Zeeman field, the resulting Hamiltonian belongs to symmetry class D rather than the usual class AII~\cite{AltlandZirnbauer1997, Schnyder2008}. We therefore use the same topological invariant, $\nu = \mathrm{sign}\det(r)$, and plot $\det(r)$ as a function of the chemical potential $\mu$, where $\mu = M - 4B$, comparing two cases: one without a disclination and one with a disclination, as shown in Fig.~\ref{fig:disclination_invariant}. In the trivial phase, $\det(r)$ remains fixed at $+1$, whereas in the GPT phase it changes sign and takes the value $-1$ over a finite parameter range, signaling a non-trivial topological phase.

\section{Method Of Construction in 3D}
\label{sec:method3d}

\subsection{Torus Geometry}
\begin{figure}[tb]
\centering
\includegraphics[width=0.48\columnwidth, height=0.4\columnwidth]{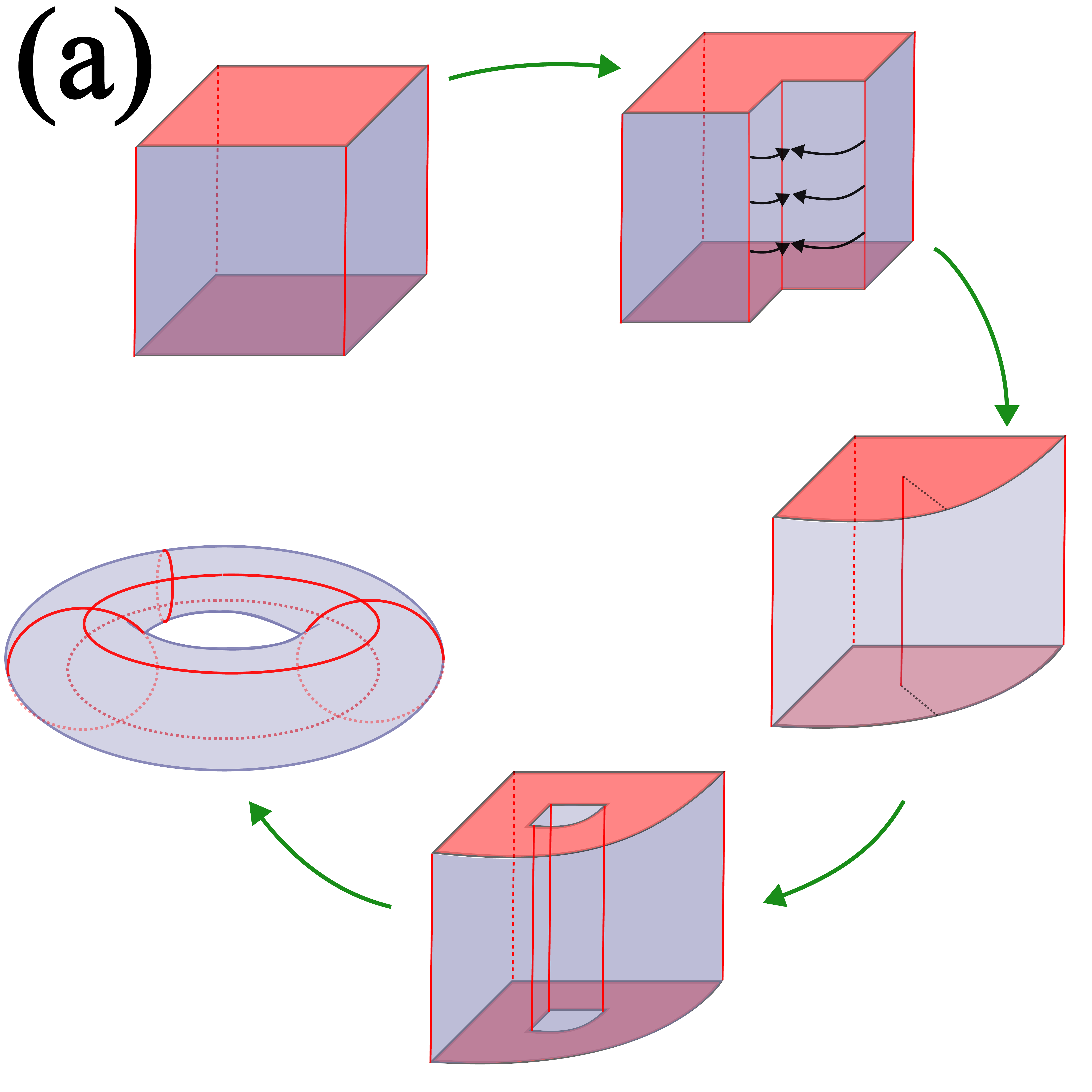}
\includegraphics[width=0.5\columnwidth, height=0.4\columnwidth]{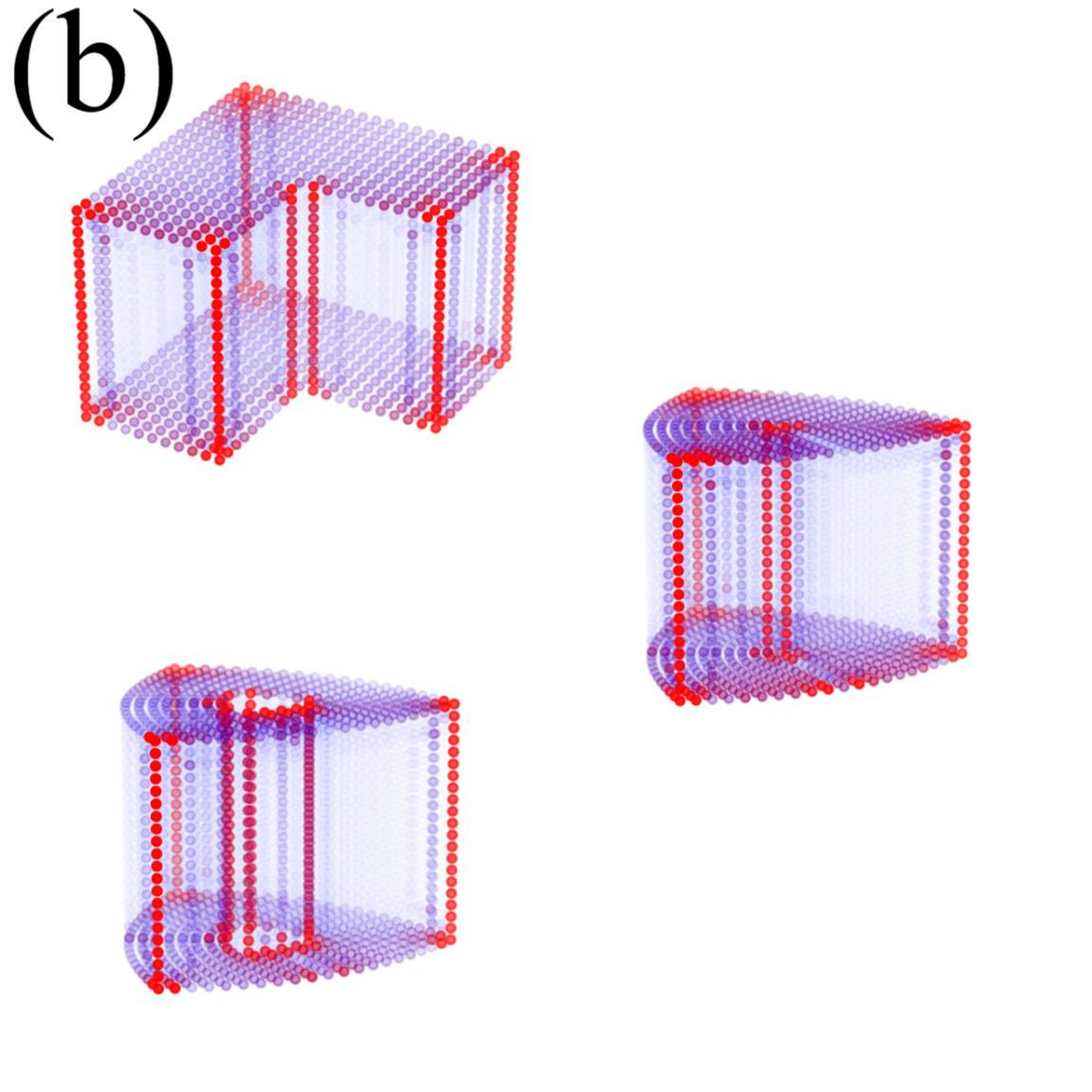}
\caption{(a) Volterra construction of a solid toroidal geometry from a cubic lattice with a $\pi/2$ disclination. The top-left figure shows the cube without a disclination, featuring gapless hinge and surface states (red). We then cut and glue edges, introduce a central hole, and round the top and bottom surfaces to form the toroidal geometry, which supports five gapless 1D helical states. (b) Real-space probability density of the low energy modes of the Hamiltonian Eq.~\eqref{eq:DIII_3D} on a cubic geometry of size $(L,l) = (20,5)$ and parameter $\delta t = 1.0t$.}
\label{fig:cube_ti_conversion}
\end{figure}

In this subsection, we outline the method of construction of GPT phases in 3D on a toroidal geometry. We begin with the class D Hamiltonian Eq.~\eqref{eq:bbh_hamiltonian} introduced in Sec.~\ref{sec:bbh_model}. To extend this model to three dimensions \cite{TeoKane2010, ShiozakiSato2014, ShiozakiSatoGomi2017}, we stack the two-dimensional layers along the $z$-axis and add a modulation dependent on $k_z$. This yields \cite{Geier_2021}
\begin{align}
H_{\text{DIII}}(\mathbf{k}) 
&= H_{\text{D}}\big[k_x, k_y;\, \delta t \to \delta t \cos(k_z)\big]\,\sigma_3  \nonumber \\
&\quad + \sin(k_z)\,\tau_0 \rho_0 \sigma_1,
\label{eq:DIII_3D}
\end{align}
where an additional set of Pauli matrices $\sigma_i$ has been introduced. The Hamiltonian $H_{\text{DIII}}(\mathbf{k})$ possesses PHS and TRS, realized by $\mathcal{K}$ and $i\tau_0\rho_0\sigma_2\mathcal{K}$, respectively. In the nontrivial phases, the system hosts helical Majorana modes protected by the underlying TRS and PHS. For a cubic geometry, as illustrated in Fig.~\ref{fig:cube_ti_conversion}, The system supports helical hinge modes propagating along the $z$ direction, while the top and bottom surfaces remain gapless and the bulk is fully gapped. The system is in a trivial GPT phase as a purely-surface perturbation can cause the helical modes propagating on the surfaces to be annihilated pairwise.

To construct the GPT phase, we follow a the Volterra construction outlined in Sec.~\ref{sec:disclinations} and introduce a $\pi/2$ disclination.
A hole is then carved into the center of the cube, and the top and bottom surfaces are subsequently rounded such as to form a toroidal geometry, as illustrated in Fig.~\ref{fig:cube_ti_conversion}(a).
When the flat top and bottom surfaces are rounded by removing lattice sites, the region where the gapless surface states are present shrinks to a ring located along the line where the surface normal remains vertical.
The resulting lattice contains five loops of helical modes, two in horizontal planes on the top and bottom, and three in vertical planes.
The two horizontal loops, and one pair of vertical loops can be continuously moved together and pairwise annihilated by surface perturbations, leaving a single noncontractible loop analogous to the one depicted in Fig.~\ref{fig:corbino_cartoon}(d). The system thus realizes a GPT phase, in which the propagating helical modes become genus-protected: they cannot be annihilated pairwise or continuously deformed to a point without either breaking the underlying TRS or traversing the bulk of the torus. We numerically simulate the geometries shown in Fig.~\ref{fig:cube_ti_conversion}(a) on a cube of length $L$, with the inner hole of square cross-section $l$ and plot the probability density of the low-energy modes in Fig.~\ref{fig:cube_ti_conversion}(b).

\section{Topological Classification}
\label{sec:topological_classification}

In this section, we discuss the topological classification of the GPT phases introduced in the previous sections. Since the number of distinct topological phases depends on the genus/number of holes of the lattice, we aim to address the following question: for a two- or three-dimensional system with genus/number of holes equal to $g$, and with the set of fundamental symmetries preserved, what is the total number of distinct GPT phases that cannot be continuously deformed into one another by purely boundary perturbations without closing the bulk gap?

\begin{figure}[tb]
\centering
\includegraphics[width=\columnwidth]{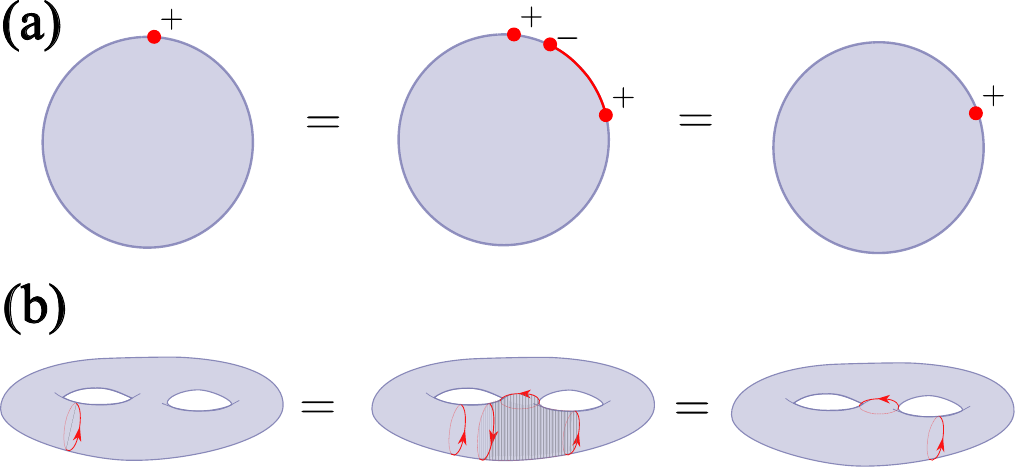}
\caption{Examples of surface perturbations in 2D (panel a) and in 3D (panel b). The initial configuration of gapless boundary states (left) can be altered by first creating a topologically-nontrivial boundary region (middle), followed by annihilating the original gapless modes with newly formed ones (right). Since this is a purely-surface perturbation, it does not change the GPT phase.}
\label{fig:moving_modes}
\end{figure}

\begin{figure}[tb]
\centering
\includegraphics[width=0.49\columnwidth, height=0.4\columnwidth]{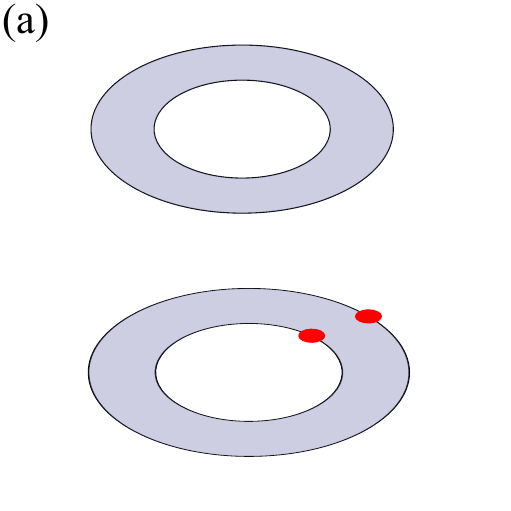}
\includegraphics[width=0.49\columnwidth, height=0.4\columnwidth]{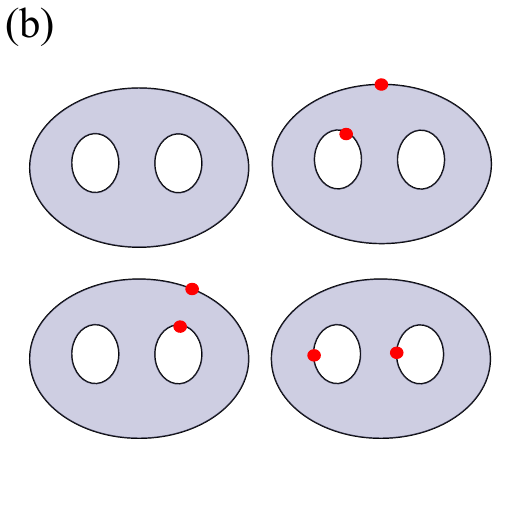}
\includegraphics[width=0.49\columnwidth, height=0.4\columnwidth]{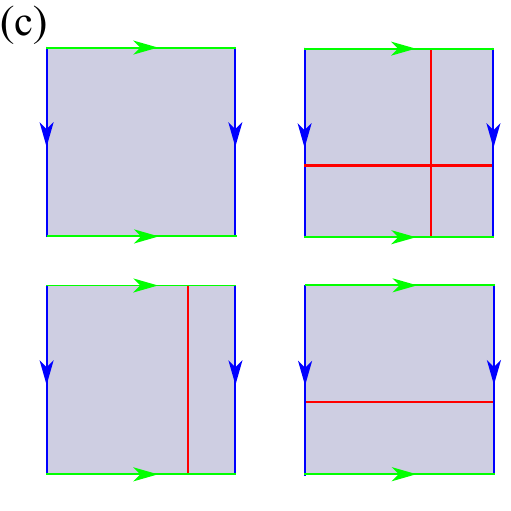}
\includegraphics[width=0.49\columnwidth, height=0.4\columnwidth]{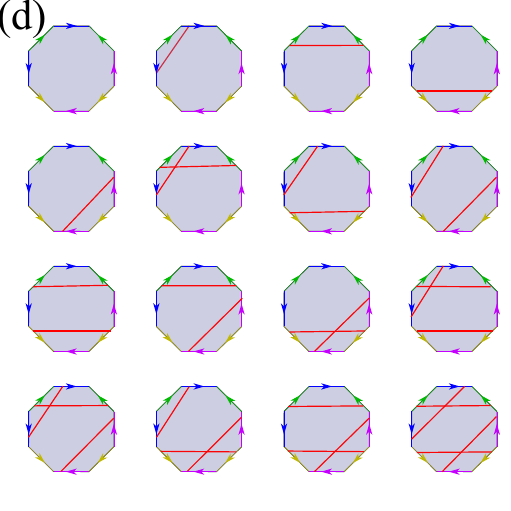}
\caption{Classification of GPT phases in two and three dimensions. (a) Two distinct GPT phases protected by PHS for a 2D Corbino-disk geometry ($g=1$), classified by $\mathbb{Z}_2$. The top panel shows the trivial phase with no zero modes, while the bottom panel shows the nontrivial phase with MZMs (red dots) localized on the edges. (b) Four distinct PHS-protected GPT phases for a 2D geometry with two holes ($g=2$), classified by $\mathbb{Z}_2^2$. 
(c) Four distinct GPT phases for a three-dimensional torus ($g=1$), classified by $\mathbb{Z}_2^{2}$. The torus surface is represented as a square with edges of the same color identified, arrows indicate the orientation of gluing, and red lines denote noncontractible loops supporting helical modes. (d) Sixteen distinct GPT phases for a three-dimensional double-torus ($g=2$), classified by $\mathbb{Z}_2^{4}$. The surface is represented as an octagon with edges of the same color identified}
\label{fig:classification_examples}
\end{figure}

We begin with two-dimensional systems and then extend the discussion to three-dimensional ones. Since the 2D GPT phases host protected boundary modes localized on one-dimensional edges, the relevant classification is given by the one-dimensional Altland-Zirnbauer (AZ) symmetry classes \cite{AltlandZirnbauer1997}. For topological systems in the AZ classes AIII, BDI, and CII, the classification yields an integer invariant $\mathbb{Z}$, whereas in classes D and DIII it yields a $\mathbb{Z}_2$ invariant. For a two-dimensional lattice with $g$ holes, there are $g+1$ disconnected edges, one of which forms the outer boundary of the lattice. The number of distinct phases can then be obtained using a simple counting argument.
Each hole can independently host an integer or $\mathbb{Z}_2$ number of modes, while the number of modes on the outer edge is fixed by the 0D topological invariant of the whole finite system (in chiral classes the net chirality of all zero modes is fixed by the trace of the chiral operator, and in class D the parity of Majorana zero modes is fixed by the total number of degrees of freedom).
A localized mode on a given edge can be moved by a surface perturbation creating a 1D topological phase on an interval that ends at the mode.
The resulting topological modes on one edge of the interval can annihilate with the mode, while the mode on the other end remains, as the net topological charge of localized modes is unchanged in this process, as shown in Fig.~\ref{fig:moving_modes}(a).
In a $\mathbb{Z}_2$ system there can be at most one protected mode per hole, so, the total number of distinct phases is $2^{g}$, and the full classification is given by $\mathbb{Z}_2^{g}$. In contrast, for $\mathbb{Z}$ systems, each hole can support an arbitrary number of modes, leading to an infinite number of distinct phases classified by $\mathbb{Z}^{g}$.
We provide examples of this classification in Fig.~\ref{fig:classification_examples}(a,b), which shows two distinct PHS-protected phases for a two-dimensional lattice with $g=1$ and four distinct PHS-protected phases for $g=2$. These phases belong to the $\mathbb{Z}_2^{g}$ classification.

For three-dimensional systems with second-order topology, an analogous argument can be applied to 1D modes on the boundary.
The boundary of a 3D system consists of a collection of orientable 2D manifolds $\Sigma_i$, each of which has a genus $g_i$.
Two configurations of 1D modes on a surface are related by a purely surface perturbation, if it is possible to create a connected region of a 2D topological phase, such that the 1D modes in the first configuration are annihilated by the new edge modes, and the remaining modes give the second configuration.
We illustrate an example of such a process in Fig.~\ref{fig:moving_modes}(b).
Protected mode configurations are exactly those, which do not arise as the boundary of a connected topological region on the surface.
We classify these configurations using homology theory \cite{Hatcher2002}.
The classification of 1D modes on a given surface $\Sigma_i$ is given by the first homology group $H_1(\Sigma_i; G)$, which classifies closed one-cycles modulo boundaries with coefficients in the group $G$.
The choice of coefficient group is determined by the topological classification of 2D phases in the given AZ symmetry class, either $\mathbb{Z}$ or $\mathbb{Z}_2$.
For $\mathbb{Z}$ classes, the 1D modes are chiral, and the invariant is integer-valued, corresponding to
\begin{equation}\label{eq:H1_Z}
H_1(\Sigma_i; \mathbb{Z}) \cong \mathbb{Z}^{2g_i}.
\end{equation}
For $\mathbb{Z}_2$ classes the modes are helical, where only the parity of modes is topologically protected, the appropriate classification is
\begin{equation}\label{eq:H1_Z2}
H_1(\Sigma_i; \mathbb{Z}_2) \cong \mathbb{Z}_2^{2g_i}.
\end{equation}
As opposed to the 2D case with Majorana modes, we do not get additional constraints from global triviality of the finite system,
hence the full classification is given by
\begin{equation}\label{eq:prod_H1}
\prod_i H_1(\Sigma_i; G) \cong G^{2\sum_i g_i},
\end{equation}
where $G$ is the 2D topological classification of the symmetry group, either $\mathbb{Z}$ or $\mathbb{Z}_2$.

To understand these homology groups in simple terms, one can think of the total number of independent, noncontractible loops that can be drawn on a closed surface. For a surface of genus $g$, there are exactly $2g$ such loops \cite{Hatcher2002}. Therefore, in the case of $\mathbb{Z}_2$ 3D HOTP, each noncontractible loop can either host a helical mode or not, leading to a $\mathbb{Z}_2^{2g}$ classification as in Eq.~\eqref{eq:H1_Z2}. For 3D HOTPs hosting chiral modes, each independent noncontractible loop can host an arbitrary number of topologically protected gapless modes, yielding $\mathbb{Z}^{2g}$ as in Eq.~\eqref{eq:H1_Z}.

We provide two concrete examples of the 3D classification in 
Fig.~\ref{fig:classification_examples}(c,d), for a system hosting helical modes, whose phases therefore belong to the $\mathbb{Z}_2^{2g}$ classification. Panel~(c) illustrates the case of a torus ($g=1$), whose surface can be represented as a square with opposite edges identified. The two independent noncontractible loops give $\mathbb{Z}_2^{2}$, giving four distinct GPT phases. Panel~(d) shows the construction for a double-torus ($g=2$), whose surface is represented as an octagon with edges identified. The four independent noncontractible loops produce $\mathbb{Z}_2^{4}$, corresponding to sixteen distinct GPT phases.

Finally we note that the 2D case can also be understood in terms of homology groups.
Here the relevant objects are closed zero-cycles modulo boundaries in the 1D boundary of the system with coefficients in the group $G$ corresponding to the 1D topological classification.
On a given boundary component $\Sigma_i$ these are classified by the zeroth homology group $H_0(\Sigma_i; G) = G$, which follows from the fact that every boundary component of a finite 2D system is a circle $S^1$.

\section{Conclusion}
\label{sec:conclusion}

In this work, we introduced genus-protected topological (GPT) phases, a class of higher-order topological phases whose gapless boundary modes are protected by the bulk gap, fundamental symmetries, and the global topology (genus/number of holes) of the system geometry, without requiring any crystalline symmetry. GPT phases share features with both intrinsic and extrinsic higher-order topological phases. These phases are realized on lattice geometries with nonzero genus/number of holes and the resulting boundary states cannot be removed by any purely surface perturbation as long as the fundamental local symmetries are preserved.

We presented construction schemes of GPT phases in two and three dimensions. In 2D, we constructed them using local edge defects introduced into the Benalcazar--Bernevig--Hughes (BBH) model on a Corbino-disk geometry, and using global topological defects, namely dislocations and disclinations, generated via a Volterra construction. In 3D, we constructed the class-DIII Hamiltonian on a toroidal geometry which yields gapless, helical 1D propagating modes on noncontractible loops of the torus.
These topological modes cannot be annihilated without either breaking the underlying symmetries or passing through the bulk. Using the $\texttt{Kwant}$ package, we plotted the probability density of the low energy modes and computed the topological index of the resulting phases by computing scattering-matrix invariants, hence proving that these systems indeed have a non-trivial topological index.

We derived the topological classification of GPT phases using a counting argument in 2D and homology theory in 3D. For a two-dimensional lattice with $g$ holes, topological phases are classified by $\mathbb{Z}_2^{g}$ or by $\mathbb{Z}^{g}$, depending on the fundamental symmetries. For a 3D system whose surface is a closed orientable two-manifold of genus $g$, helical modes are classified by $\mathbb{Z}_2^{2g}$ while chiral modes by $\mathbb{Z}^{2g}$.

Finally, we believe that the GPT phases introduced in this work are experimentally accessible and can be realized in several platforms. 
\textit{Topolectrical circuits}~\cite{Lee2018Topoelectrical,Imhof2018Topoelectrical}, which implement tight-binding Hamiltonians using networks of inductors, capacitors, and resistors, provide a particularly natural setting. In such systems the genus of the lattice is determined entirely by the connectivity of the circuit graph rather than by the physical shape of the network, making it straightforward to engineer geometries with nontrivial genus. Higher-order topological phases have already been realized in topolectrical circuits~\cite{Imhof2018Topoelectrical,Bao2019Topoelectrical,Lv2021Topoelectrical,Dong2021Topolectrical}, and lattice defects such as dislocations and disclinations employed in our construction can be introduced \cite{Peterson2021Topoelectrical,Yamada2022Topoelectrical}. 
\textit{Photonic platforms} provide another promising route for realizing GPT phases. Higher-order topological phases have been observed in photonic crystals and waveguide arrays~\cite{Chen2019Photonic,Xie2019Photonic,Li2020Photonic}, and photonic systems also allow the introduction of lattice defects such as dislocations and disclinations~\cite{Lin2023Photonic,Lustig2022Photonic,Liu2021Photonic}. Beyond these platforms, GPT phases might also be implemented in other metamaterial systems, including acoustic and mechanical lattices \cite{SerraGarcia2018MetaMaterial, Xue2019MetaMaterial,Ye2022Acoustic,Deng2022Acoustic}. 

\section{Acknowledgments}
We thank Ulrike Nitzsche for technical assistance.
We are grateful for discussions on homology theory to Gerg\H{o} Pint\'{e}r.
This work was supported by the Deutsche Forschungsgemeinschaft (DFG, German Research Foundation) under Germany’s Excellence Strategy through the W\"{u}rzburg-Dresden Cluster of Excellence on Complexity and Topology in Quantum Matter – ctd.qmat (EXC 2147, 390858490 and 392019).
D.~V. was supported by the National Research, Development and Innovation Office of Hungary under OTKA grant no. FK 146499, and the János Bolyai Research Scholarship of the Hungarian Academy of Sciences.

\bibliography{refs}

\end{document}